\documentclass[aps,prl,twocolumn,superscriptaddress,showpacs]{revtex4}
\usepackage{graphicx,graphics,amssymb,amsmath}
\usepackage[usenames,dvipsnames]{color}
\begin{document}

%\preprint{}

\title{Microscopic properties of the ``pinwheel" kagome compound Rb$_2$Cu$_3$SnF$_{12}$}
%\title{\textcolor{red}{Spin gap and transverse moments} in the ``pinwheel'' kagome Rb$_2$Cu$_3$SnF$_{12}$ \textcolor{red}{under high magnetic field}}


\author{M. S. Grbi\'c}
\email[]{mgrbic@phy.hr}
%\thanks{}
\affiliation{Laboratoire National des Champs Magn\' etiques Intenses, LNCMI - CNRS (UPR3228), UJF, UPS and INSA, BP 166, 38042 Grenoble Cedex 9, France}
\affiliation{Department of Physics, Faculty of Science, University of Zagreb, P.O. Box 331, HR-10002 Zagreb, Croatia}

\author{S. Kr\" amer}
%\thanks{}
\affiliation{Laboratoire National des Champs Magn\' etiques Intenses, LNCMI - CNRS (UPR3228), UJF, UPS and INSA, BP 166, 38042 Grenoble Cedex 9, France}

\author{C. Berthier}
%\thanks{}
\affiliation{Laboratoire National des Champs Magn\' etiques Intenses, LNCMI - CNRS (UPR3228), UJF, UPS and INSA, BP 166, 38042 Grenoble Cedex 9, France}

\author{F. Trousselet}
%\thanks{}
\affiliation{Institut N\' eel, CNRS and Universit\' e Joseph Fourier, BP 166, 38042 Grenoble Cedex 9, France}

\author{O. C\' epas}
%\thanks{}
\affiliation{Institut N\' eel, CNRS and Universit\' e Joseph Fourier, BP 166, 38042 Grenoble Cedex 9, France}

\author{H. Tanaka}
%\thanks{}
\affiliation{Department of Physics, Tokyo Institute of Technology, Meguro-ku, Tokyo 152-8551, Japan}

\author{M. Horvati\' c}
\email[]{mladen.horvatic@lncmi.cnrs.fr}
%\thanks{}
\affiliation{Laboratoire National des Champs Magn\' etiques Intenses, LNCMI - CNRS (UPR3228), UJF, UPS and INSA, BP 166, 38042 Grenoble Cedex 9, France}            
\date{\today}

\begin{abstract}
\indent Using $^{63,65}$Cu nuclear magnetic resonance (NMR) in magnetic fields up to 30 T we study the microscopic properties of the 12-site valence-bond-solid ground state in the ``pinwheel" kagome compound Rb$_2$Cu$_3$SnF$_{12}$. We find that the ground state is characterized by a strong transverse staggered spin polarization whose temperature and field dependence points to a mixing of the singlet and triplet states. This is further corroborated by the field dependence of the gap $\Delta (H)$, which has a level anticrossing with a large minimum gap value of $\approx \Delta (0)/2$, with no evidence of a phase transition down to 1.5\,K. By the exact diagonalization of small clusters, we show that the observed anticrossing is mainly due to staggered tilts of the $g$-tensors defined by the crystal structure, and reveal symmetry properties of the low-energy excitation spectrum compatible with the absence of level crossing.

%We study the high magnetic field properties, up to 30 T, of the spin-gapped valence-bond-solid
%state of the 2d ``pinwheel'' kagome compound Rb$_2$Cu$_3$SnF$_{12}$,
%by $\strut^{63,65}{\rm Cu}$ nuclear magnetic resonance (NMR).
%In contrast to the expected field-induced phase transition}, we
%find a spin gap showing a finite minimum at 14 T and
%large transverse staggered moments in the ground state. We explain that these
%effects are primarily due to the staggered $g$-tensors. They generate
%intrinsic transverse moments, which break the same symmetries as the
%ones spontaneously broken in the phase transition.  The
%transition is therefore predicted to be \textit{rounded} and, indeed, no sharp phase
%transition was found, neither in field nor in temperature down to 1.5~K.
\end{abstract}

%\keywords{}

\maketitle

%\section{}

\maketitle

%\section{Model}

\begin{figure*}[t!]
\includegraphics[width=\textwidth]{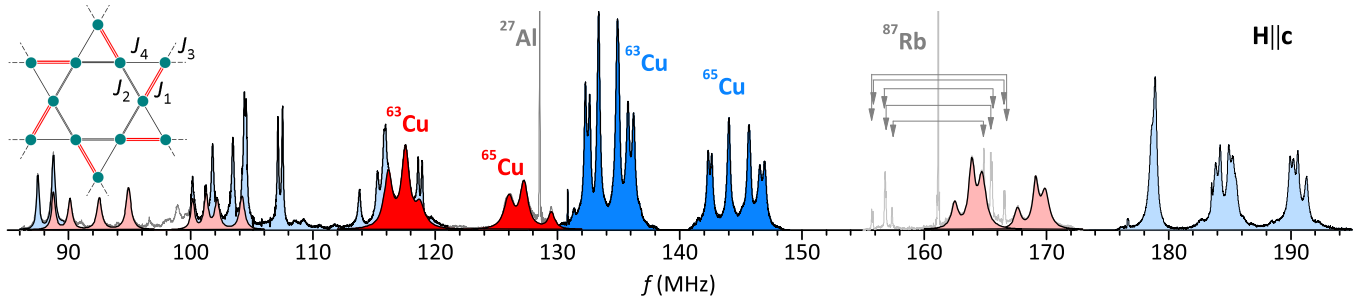}
\caption{\label{spectrum}(color online) $^{63,65}$Cu spectrum of Rb$_2$Cu$_3$SnF$_{12}$ measured at 2.6\,K and 11.568\,T for the magnetic field parallel to the $c$-axis of the crystal. Colors and their hues are used to distinguish positively (red) and negatively (blue) polarized sites, as well as the central lines (dark color) and satellites (light color). Additional lines of $^{87}$Rb and $^{27}$Al (used for magnetic field calibration) are marked in grey. The inset shows the exchange couplings structure of a pinwheel cluster.
}
\end{figure*}

\begin{figure}[t!]
\includegraphics[width=\columnwidth]{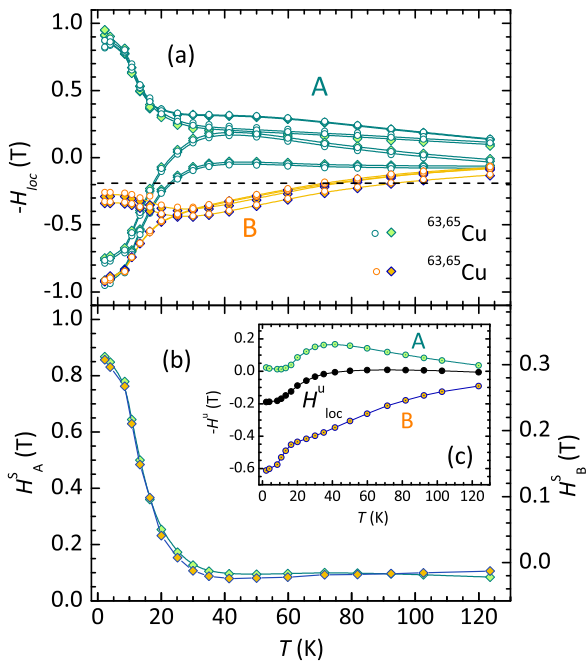}
\caption{(color online) (a) Temperature dependence of the local fields at 13.0 T. Two groups of sites are marked by the green (A) and orange (B) symbols. Empty (filled) symbols show the values obtained for $^{63}$Cu ($^{65}$Cu) isotope. %while the lines are guide to the eye passing through their average value.
The dashed line marks estimated orbital contribution ($K_{\rm{orb}}~\approx~1.5$\,\%). (b) The average local {\it{staggered}} field at sites A (green, left scale) and B (orange, right scale). (c) The {\it{uniform}} hyperfine field averaged over the A, B and all sites (green, orange and black circles, respectively).}
\label{Spread}
\end{figure}

\begin{figure}[t!]
\includegraphics[width=\columnwidth]{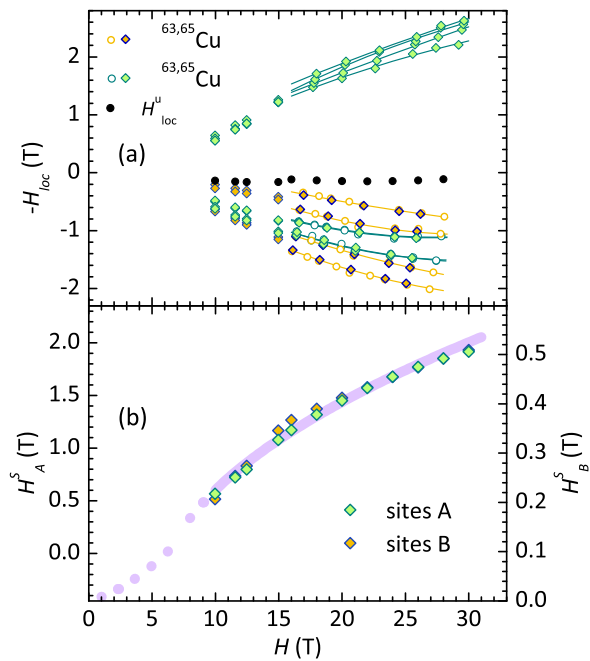}
\caption{(color online) (a) Magnetic field dependence of the local fields at 3.9~K, with the same symbol and color code as in Fig.~\ref{Spread}. (b) The average local staggered field. The thick line is a guide for the eye showing a square root dependence, while the dotted line shows an expected qualitative behavior for lower magnetic fields.}
\label{SpreadH}%
\end{figure}

\begin{figure}[t!]%
\includegraphics[width=\columnwidth]{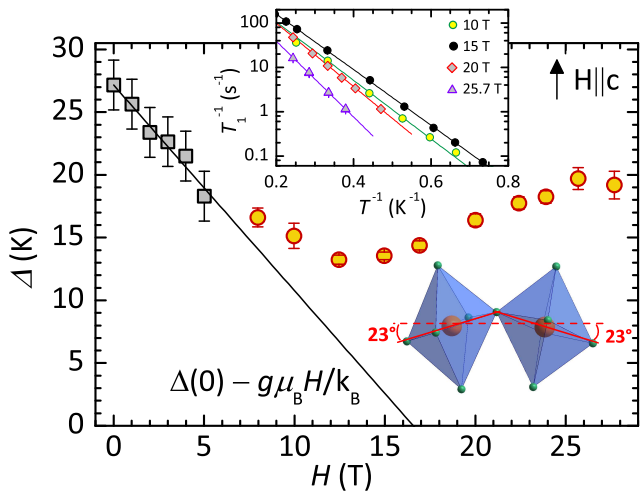}%
\caption{(color online) $H$- dependence of the singlet-triplet energy gap measured by $T^{-1} _1$ (red-yellow circles) and by neutron scattering (black-grey squares, from Ref.~\cite{Matan}). Insets: (top) Arrhenius plot of the measured $T_1^{-1}(T)$ dependence (symbols) and the fit (lines) used to extract the gap values shown in the main figure. Only selected, representative data sets are shown. (bottom-right) Two neighboring CuF$_6$ octahedra.}%
\label{Gap}%
\end{figure}

Frustrated magnetic systems attract a lot of interest due to the interplay
of frustration and quantum effects which bring about the emergence
of many exotic ground states and fractionalized excitations~\cite{Normand,Balents}. The ($S$\,=\,1/2)
Heisenberg model on the kagome lattice has a particularly high level of frustration
leading to a nonmagnetic ground state and many competing singlet states at low energy ~\cite{Lecheminant}.
%\sout{It is well established that a large number of singlet states,
%  the number growing exponentially with the size of the system, is
%  lying below the triplet states within a spin gap
%  $\Delta$\,$\sim$\,$J/20$ (where $J$ is the strength of the
%  antiferromagnetic coupling)~\cite{Lecheminant}. However, despite
%  many theoretical efforts, the nature of the ground state remains an
%  open question. There have been different proposals for the ground
%  state,} 
The nature of the ground state is currently highly debated~\cite{Yan}. Proposals include various quantum spin liquids
%(\sout{~\cite{Anderson,Sachdev,Ran07,Ran09}})
%\textcolor{blue}{[My point of view: either put all refs, none, or only those relevant to the present work, see below Syromyatnikov and Capponi. Note also
%    that Ref. Ran09 deals with spontaneous ordering when the field is
%    applied]}
with unbroken lattice symmetry and valence bond solids (VBS)
%\sout{~\cite{Marston,Yang08,Singh}}
where the lattice symmetry is
broken. Unfortunately, only a handful of real systems do not
magnetically order at low temperatures, and not all of them are well suited to
study the properties of their ground states. Even fewer can be
grown as single crystals and without the intrinsic disorder which complicates the
interpretation of experimental results.\\
\indent Recently, a singlet ground state has been found in the distorted kagome system
Rb$_2$Cu$_3$SnF$_{12}$ ~\cite{Morita}, where the $S$\,=\,1/2 spin is
carried by the copper Cu$^{2+}$ ion enclosed in an F$_6$
octahedron. The distortion of the kagome lattice can be seen as 6
elongated hexagons surrounding a regular one, and the spins are
connected through the Cu$^{2+}-$F$^--$Cu$^{2+}$ bonds with 4 different
bonding angles, creating 4 different exchange couplings. Magnetization
measurements at high magnetic fields showed a crossover in the behavior
of the system appearing between 10 and 20~T for the magnetic field
perpendicular to the kagome planes (\textbf{H}~$\parallel$~$c$~axis)
and a gradual filling of the triplet band. It was
proposed~\cite{Yang09} that the strongest coupling ($J_1$~=~216\,K~\cite{Matan})
creates valence bonds between spin pairs and forms a ``pinwheel"
pattern around the regular kagome hexagon, with twelve sites in the unit-cell (inset to
Fig.~\ref{spectrum}). The remaining three couplings ($J_2 = 0.95J_1$,
$J_3 = 0.85J_1$, $J_4 = 0.55J_1$) are somewhat weaker,
%\sout{and an additional Dzyaloshinskii-Moriya (DM) interaction further
%  define the energy structure of the system. A 12-site VBS state was
%  shown to be energetically preferable to the 36-site VBS state
%  proposed for the uniform kagome antiferromagnet.}
 and what remains of the pure kagome physics is an open question. It is important to note that, for the pure kagome model, a VBS state with a 12-site unit cell was originally described as resonating between the two ``pinwheel'' patterns~\cite{Syromyatnikov}. It was argued that the distortions observed in Rb$_2$Cu$_3$SnF$_{12}$ stabilize the VBS with a single ``pinwheel'' pattern~\cite{Yang09}, and such a chirality breaking has been recently predicted to occur spontaneously in the pure kagome model~\cite{Capponi}. A neutron scattering study~\cite{Matan} has found evidence of a 12-site VBS ground state, separated from the first triplet state with strongly renormalized energy gap $\Delta(H$\,=\,0) = 27\,K $\approx J_1/8$. Measurements in magnetic field up to 6\,T applied along the $c$ axis (perpendicular to the kagome planes) show a reduction of the singlet-triplet gap and lifting of the triplet band degeneracy, accounted for by a \textit{longitudinal} Dzyaloshinskii-Moriya (DM) interaction (\textbf{D} $\parallel$~$c$)~\cite{Matan} which does not break the rotational U(1) symmetry.\\
\indent The availability of Rb$_2$Cu$_3$SnF$_{12}$ in the form of large single
crystals has already stimulated fruitful experimental and theoretical
work~\cite{Morita,Yang09,Matan,Hwang,Khatami}. However, the
microscopic properties of the 12-site ``pinwheel" VBS ground state, as
well as its behavior at high magnetic field, are still unknown. In
this Letter we report an NMR study of the on-site copper $^{63,65}$Cu
nuclei in magnetic fields up to 30\,T, applied parallel to the
$c$ axis (experimental details can be found in~\cite{CF}). We find evidence of an unconventional magnetic lattice with
strong staggered \textit{transverse} magnetic moments. We determine the field
dependence of the singlet-triplet gap, and show that there is no phase
transition connected with the closing of the gap, but rather a
level anticrossing that keeps the gap open. Together
 with the field dependence of local spin polarizations, this points to a mixing
of the singlet and triplet states, typically associated with %with U(1)-symmetry breaking DM interaction \textit{perpendicular} to the field and/or a staggered off-diagonal $g$-tensor component $g_s$.
DM interaction \textit{perpendicular} to the field and/or a staggered off-diagonal $g$-tensor component $g_s$, that both break the U(1) rotational symmetry.
% These properties provide a very stringent framework for a theoretical description of Rb$_2$Cu$_3$SnF$_{12}$ compound, and 
%By using exact diagonalization on small clusters we show that the very existence of anticrossing defines a symmetry condition on the U(1)-symmetry breaking interaction, that the anticrossing gap is dominantly due to $g_s$, and that their symmetry thus define the sign of the longitudinal DM interaction.\\
%By using exact diagonalization on small clusters we show that the very existence of the anticrossing imposes an additional condition on these \textit{transverse} anisotropies, allowing us to conclude that the anticrossing gap is dominantly due to a specific symmetry of $g_s$ which defines the sign of the longitudinal DM interaction.\\
We argue that the very existence of the anticrossing imposes an additional symmetry condition on the lowest triplet state, which defines the sign of the longitudinal DM interaction. By using exact diagonalization of small clusters,  we show that the amplitude of the anticrossing gap is dominantly due to $g_s$ terms.\\ 
\indent When a single crystal sample is placed in a magnetic field, the NMR spectrum of the copper $^{63,65}$Cu (spin 3/2) nuclei will show 3 lines per isotope for each non-equivalent site: one central line surrounded by two satellites. This sextuplet of lines is multiplied by the number of inequivalent sites, e.g., the spectrum shown in Fig.~\ref{spectrum} presents a very complicated structure spanning over 100 MHz, consisting of 72 NMR lines originating from 12 sites. To analyse such spectra, we follow closely the procedure and nomenclature used in Ref.~\cite{Aimo}, where the technical details are in the second column of page 2. The six NMR frequencies of each sextuplet of lines belonging to one copper site are determined by the quadrupolar coupling tensor, described by its principal component $\nu_Q$ and the asymmetry parameter $\eta$, by the Zeeman coupling to the local magnetic field $H_{\rm{eff}}$, and by the ($\vartheta$,\,$\varphi$) angles defining relative direction of that field with respect to the principal axes of the quadrupolar tensor. Here, these 5 parameters are further constrained by the known nuclear quadrupolar frequency  $\nu_{NQR} =  \nu_Q \sqrt{1+ \eta^2/3}$, as four different $\nu_{NQR}$ frequencies, all in the narrow range 49.8-55.3~MHz, have been reported in a previous study~\cite{Tashiro}. The main difficulty in the analysis of the spectra lies in the assignments of different sextuplets to lines in the NMR spectrum. This requires numerous trials and errors, where the correct assignment was recognized by successful fit to several complete spectra taken at different field values, leading to reasonable values of the fit parameters ($H_{\rm{eff}}$,~$\vartheta$,~$\varphi$,~$\nu_Q$,~$\eta$)$_i$ , where $i$ denotes different Cu sites. In particular, for nearly all the sites we found $\vartheta \approx 24$\,$\pm$\,2$^\circ$, (while only two sites have somewhat smaller values $13$\,$\pm$\,2$^\circ$) which corresponds precisely to what is expected from the crystal structure~\cite{Morita}. That is, for all CuF$_6$ octahedra, the (average) planes passing through 4 tetragonally placed F$^-$ ions that define the $d_{x^2-y^2}$ orbital of copper are tilted by $22^\circ$-$23^\circ$ with respect to the kagome plane (see inset to Fig.~\ref{Gap}).\\
\indent Among the complete set of parameters, only the $H_{\rm{eff}}$ values are expected to be temperature ($T$) and field dependent. Once determined, the other parameters can be taken as constants, so that in the study of $T$- and $H$-dependence of spectra, from each NMR line position we can calculate the corresponding $H_{\rm{eff}}$ value. For these studies we have thus measured only the central lines of the spectra, meaning that each Cu site was represented by two lines from two $^{65,63}$Cu isotopes which should lead to the same $H_{\rm{eff}}$ value. As shown in Figs.~\ref{Spread}(a) and~\ref{SpreadH}(a) this is indeed the case. In these figures we have plotted the negative of the local field $-H_{\rm{loc}} = H - H_{\rm{eff}}$, where $H$ is the applied field, which is directly proportional to the local spin polarization, $-H_{\rm{loc}} =  -\textbf{A} \times 2 \left\langle \textbf{S}\right\rangle - \textbf{K}_{\rm{orb}} \textbf{H}$, up to relatively small constant correction due to the orbital contribution, estimated as $K_{\rm{orb}} \approx 1.5$\,\%. Typical values of the hyperfine coupling tensor \textbf{A} for the Cu nuclei in the CuF$_6$ environment are known~\cite{Kubo}, $A_{\parallel} \approx -18$\,T, $A_{\parallel} /A_{\bot}  \approx$~10, where ``$\parallel$" denotes the principal axis of the \textbf{A} tensor, expected to be nearly parallel to the principal axis of the quadrupolar coupling tensor, i.e. tilted by $\approx$$\vartheta$ from \textbf{H} $\parallel c$. In this case, both the longitudinal coupling constant $A_{zz} = A_{\parallel}$cos$^2\vartheta + A_{\bot} $sin$^2\vartheta$, as well as the transverse one $A_{z \bot}  = \frac{1}{2}(A_{\parallel} - A_{\bot})$\,sin\,$2\vartheta$\,cos\,$\phi$ are known, and the measured $-H_{\rm{loc}}$ directly reflects the corresponding local spin polarization ($S_z$, $S_{\bot}$, $\phi$), where $\phi$ denotes the azimuthal angle relative to the one defined by the $\vartheta$-tilted $A_{\parallel}$ axis.
% with respect to the direction of the $\vartheta$ tilt. 
It is clear that from one number, $H_{\rm{loc}}$, one cannot {\it{a priori}} deduce three spin components. The necessary information to get $S_z$ and $S_{\bot}$ spin components is thus obtained from the analysis of the $-H_{\rm{loc}}(T, H)$ dependence (Figs.~\ref{Spread} and~\ref{SpreadH}).\\
\indent Our NMR spectrum shown in Fig.~\ref{spectrum} has a large number of lines, corresponding to many different local fields originating from crystallographically and/or magnetically different sites. In contrast to that, the room temperature crystal structure predicts only {\it{two}} inequivalent copper sites in the unit cell. In order to understand the origin of the observed distribution of local fields, we have tracked their temperature dependence, shown in Fig.~\ref{Spread}(a). There, one can identify two families of  sites (marked with A (green) and B (orange)) occupied in the ratio $P(A)$\,:\,$P(B)$ = 8\,:\,4 = 2\,:\,1, which can be associated with 12 sites per unit cell of the crystal structure below 215\,K. Close to this temperature a small structural distortion has been observed by X-ray scattering~\cite{Matan}, leading to the enlargement of the lattice cell to $2a$$\times$$2a$, but the superlattice peaks were too weak to fully resolve the new structure at lower $T$. While this structural transition was not detected in the dc susceptibility, such subtle deformations can cause the splitting of the NMR lines. Furthermore, in Fig.~\ref{Spread}(a) it is easy to observe the local ``pairs" of each family that develop opposite polarizations at low temperature, defining thus the local staggered ($H_{A,B}^s$) and uniform ($H_{A,B}^u$) fields,
\begin{eqnarray}\label{Eq}
     H_{A,B}^s &= (\left\langle  H_{i+} \right\rangle_{A,B} - \left\langle H_{i-}  \right\rangle_{A,B})/2 ,\nonumber \\
         H_{A,B}^u &= (\left\langle  H_{i+} \right\rangle_{A,B} + \left\langle H_{i-}\right\rangle_{A,B})/2 ,
\end{eqnarray}
where the average is taken over the upper ($H_{i+}$) or lower ($H_{i-}$) local fields of either A or B family of lines. In Fig.~\ref{Spread}(b) we can see that both A and B sites develop very strong staggered moments below 30\,K ($\approx \Delta(0)$). Additionally (see Fig.~\ref{Spread}(c)), the two families of lines also develop distinct positive ($H_A^u$) and negative ($H_B^u$) uniform magnetizations. When $T$ is raised, the distribution of the local fields is reduced but retains a finite value, even at 120\,K ($\approx 4\Delta(0)$)~\cite{comment1}. This site-inequivalence is attributed to the previously mentioned structural distortion of the lattice~\cite{Matan}. On the other hand, the development of the staggered ($H_{\rm{loc}}^s$) and uniform ($H_{\rm{loc}}^u$) fields at temperatures below the zero field gap in Fig.~\ref{Spread}(b)
is clearly related to the formation of the local spin order.\\
\indent As already mentioned, both the transverse and longitudinal magnetization contribute to $H_{\rm{loc}}$, and, due to the uncertainty in the site environments and the complexity of the spectra, we could not find a way to formally separate the two contributions. However, considering their relative size at low temperature, $\left| H^s \right| \gg \left| H^u \right|$, it is natural to associate the staggered (uniform) fields $H^{s(u)}$ defined by Eqs.~(\ref{Eq}) to the transverse (longitudinal) local spin polarization $S_{\bot} (S_z)$, respectively. This is further supported by the field dependence of these values and the physics behind them, as discussed later. Within this attribution we find very big low-$T$ values for the local transverse staggered spin polarization, $S_{\bot}$cos\,$\phi \approx$ 14\,\% and 6\,\% for the A and B sites respectively, while the local uniform fields correspond to much smaller spin polarization of $S_z \approx 1.2$\,\% and $-2$\,\%. While in this way only a projection $S_{\bot}$cos\,$\phi$ is determined, in all cases the spins polarizations lie nearly completely in the kagome plane, $S_{\bot} \gg \left| S_z \right|$.\\
\indent We observe that the total average  $\left\langle S_z \right\rangle$  component is almost fully cancelled out, $P(A)  \left\langle S_z\right\rangle_A + P(B)  \left\langle S_z\right\rangle_B \approx 0$, but the true value cannot be precisely defined, because of the uncertainty in the estimate of the orbital shift $K_{\rm{orb}}$
defining the zero. As regards the transverse staggered components, the smaller value of $H_B^s$ for sites B can be explained either as a weaker polarized moments having equivalent orientations, $\phi_A \approx \phi_B$, or moments of similar size but different $\phi$ values, e.g.,  $\phi_A \approx 0$ and  $\phi_B \approx 65^{\circ}$. We cannot differentiate between these two cases, but the existence of a large staggered magnetization within the kagome planes is not affected by this uncertainty.\\
\indent We have also followed the magnetic field dependence of the local polarizations up to 30\,T (Fig.~\ref{SpreadH}), measured at 3.9\,K where the temperature dependence of the spin polarizations saturates. The field dependence of $H_{A,B}^s$ (Fig.~\ref{SpreadH}(b)) is clearly different from the high field magnetization~\cite{Morita} and follows approximately a square-root dependence, reaching at 30\,T a polarization of $S_{\bot}$cos\,$\phi \approx 30$\,\% for the A sites. Similar temperature and magnetic field dependence of local moments was found~\cite{Kodama} in SrCu$_2$(BO$_3$)$_2$, another 2D frustrated dimer system. There, a moderate in-plane DM interaction ($D/J = 0.034$) and $g_s$ (= 0.023) terms were enough to mix the singlet and triplet states and thus create considerable transverse staggered moments, while the longitudinal moments remained much smaller. In Rb$_2$Cu$_3$SnF$_{12}$ the DM interaction is estimated to be much stronger ($D/J = 0.18-0.2$) \cite{Matan, Hwang}, but only the longitudinal $c$-axis component of the \textbf{D} vector, which does \textit{not} induce level-mixing, has been considered so far.\\
%Our results imply that a significant mixing term should also be taken into consideration.\\
%either a DM interaction perpendicular to the field ($d_p$) and/or a staggered $g$ tensor component $g_s$ should also be taken into consideration.\\
\indent While the application of the critical field in spin-dimer systems is expected to close the singlet-triplet gap by the level-crossing, leading to a new, polarized phase, mixing of the singlet and triplet states will keep the gap open (``anticrossing"). In order to follow the magnetic field dependence of this gap, we have performed measurements of the nuclear spin-lattice relaxation rate $T_{1}^{-1}$, which probes the dynamics of the low-energy excitations. Importantly, no qualitative changes in the spectra nor any $T_1^{-1}$ anomaly were observed down to 1.5\,K, confirming the absence of a phase transition. For a fixed field, the temperature dependence of $T_1^{-1}$ below 6\,K (top inset of Fig.~\ref{Gap}) shows an activated behavior, $\propto e^{-\Delta(H)/T}$, typical of a 2-magnon process, allowing us to determine the value of the gap, $\Delta(H)$~\cite{comment2}. As the field is raised from 8\,T to 13\,T, the gap value is seen to decrease (Fig.~\ref{Gap}), and the NMR data smoothly continue those determined from the neutron-scattering up to 6\,T~\cite{Matan}. Close to 13\,T the gap value passes through a broad minimum after which it again increases at higher fields. The residual value of the gap $\Delta(13~\rm{T}) \approx \Delta(0)/2$ appears to be very large, which is a clear evidence for important U(1)-symmetry breaking terms that induce level anticrossing.\\ %\textcolor{red}{transverse anisotropies}.
\indent In order to describe this anticrossing, we have included the
transverse anisotropies in the Hamiltonian, which was defined previously
by the Heisenberg couplings $J_1 - J_4$ and the $z$ component ($d_z$)
of the \textbf{D} vector~\cite{Matan, Hwang}, and performed exact diagonalizations of this Hamiltonian on clusters of $N$=~12 and 24 sites \cite{CF}. We consider the in-plane DM terms ($d_p = D_{\perp i}/J_i$) perpendicular
to the bonds and respecting the crystal symmetry, as well as the
off-diagonal $g$-tensor terms as generated by the strong tilting of
the CuF$_6$ octahedra, $\vartheta \approx 23^{\circ}$ (inset of
Fig.~\ref{Gap}), leading to $g_s = 0.135$ (see \cite{CF} for
definitions). Remarkably, depending on the helicity of these terms as compared to that defined by the screw axis of the $d_z$ vector, the system presents either level crossing or anticrossing. That is, depending on the sign of $d_z$, the lowest triplet state will be either of $E_g^{+}$ or $E_g^{-}$ symmetry, and only one of these two (in the present case $E_g^{-}$ corresponding to $d_z>0$ \cite{CF}) allows for level anticrossing. The calculated gap presents \textit{small} finite-size effects near the anticrossing, in contrast to the zero-field gap as described in~\mbox{\cite{Matan,Hwang}}. This enabled us to make a theoretical estimate of the residual gap size and position, and conclude that the experimentally observed gap is well accounted for by the $g_s$ terms alone, with calculated values $\Delta_{res}^{th}=13.0$~K and $H_c^{th} = 14.0$~T \cite{CF}, while only minor contribution can come from $d_p$ (estimated to $\left|d_p\right| < 0.012$).\\
\indent In summary, we have described the microscopic properties of Rb$_2$Cu$_3$SnF$_{12}$ by on-site $^{63,65}$Cu NMR.
In the field perpendicular to the kagome planes and at low temperature, the NMR spectra evidence a strong staggered transverse spin polarization, growing approximately as a square-root of magnetic field. 
The field dependence of the singlet-triplet gap measured via the $T_1^{-1}$ data presents an anticrossing of the energy levels with a large residual gap value $\Delta(13~\rm{T}) \approx \Delta(0)/2$, well accounted for by the staggered $g$-tensor terms defined by the crystal structure. 
% provided that $d_z<0$. The absence of phase transition in this compound is the consequence of the staggered $g$-tensors, which induce intrinsic transverse moments of the same symmetry as that which would otherwise be spontaneously broken. 
%From exact diagonalization of small clusters we learn that the existence of anticrossing fully defines the symmetry of the lowest triplet state, and is well accounted for by the staggered $g$-tensors. 
We have argued that the observed anticrossing and the absence of phase transition is only compatible with a given sign of $d_z$. The other sign would lead to a phase transition because the spontaneously induced transverse moments would break the rotation symmetry of the crystal. Further theoretical work is needed to fully exploit the available information on spin polarizations.\\
\indent We acknowledge fruitful discussions with Y. Fukumoto, H. Mayaffre and S. Maegawa. We thank W. G. Clark for reviewing the manuscript. Part of this work has been supported by the French ANR project NEMSICOM, by the EuroMagNET network under the EU contract No. 228043, by the ARRS project No. J1-2118, and by the EU FP7 project SOLeNeMaR No. 229390.

\clearpage

\section{Supplemental material}
\vspace{-0.4cm}
\subsection{to `Microscopic properties of the ``pinwheel" kagome compound Rb$_2$Cu$_3$SnF$_{12}$' by M. S. Grbi\'c \textit{et al.}}  

\renewcommand{\thefigure}{S\arabic{figure}}
\setcounter{figure}{0}
\renewcommand{\theequation}{S\arabic{equation}}
\setcounter{equation}{0}

%\subsection{Technical details of the measurements}
\subsection{Experimental methods}             
Transparent single crystals of Rb$_2$Cu$_3$SnF$_{12}$ were synthesized according to the procedure described in ref.~\cite{Morita}. The size of the crystal used for the NMR study was 3$\times$4$\times$1 mm$^3$. The sample was placed in a silver-wire radio-frequency coil together with a small piece of thin aluminum foil, which was used to calibrate the value of the applied magnetic field $H$. The $c$ axis of the crystal was oriented within 2$^\circ$ along the field, as confirmed from the angular dependence of the spectrum. A laboratory-made NMR spectrometer was used for the data acquisition; the spectra were recorded using a Hahn-echo pulse sequence, $\pi/2 - \tau - \pi$, with typical $\pi$ pulse value of 3 $\mu$s, and $\tau = 10$ $\mu$s. An inversion-recovery pulse sequence was used for measurements of the spin-lattice relaxation rate. The measurements in fields up to 17 T were made in a superconducting NMR magnet, while the data at higher field values were taken using a 20 MW resistive magnet at the LNCMI - Grenoble. 

%Transparent crystals of Rb$_2$Cu$_3$SnF$_{12}$ were synthesized according to the procedure described in ref.~\cite{Morita}. Size of the crystal used for the study was $3 \times 4 \times 1$ mm$^3$. Sample was placed in a silver single-turn rf coil, together with a small piece of thin aluminum foil that was used to precisely determine the value of the applied magnetic field $H$. The c axis of the crystal was oriented within 2$^\circ$ along $\textbf{H}$, as indicated by the angular variation of the spectrum. For the data acquisition we used a home-made NMR spectrometer and a Hahn-echo pulse sequence $\pi/2 - \tau - \pi$. Typical length of the $\pi$ pulse was 3 us, and $\tau = 10$ $\mu$s. For measurements of the spin-lattice relaxation time, we used an inversion-pulse recovery sequence. Both, spectrum and relaxation of the nucleus were determined from a Fourier-transformed spin-echo signal. Measurements in fields up to 17 T were made in an Oxford Instruments superconducting magnet, while data at higher field values were taken using a 20 MW resistive magnet at the Laboratorie National des Champs Magn\'etiqes Intenses, Grenoble (France).

\subsection{Anticrossing at high magnetic field in the pinwheel kagome Rb$_2$Cu$_3$SnF$_{12}$}

%\pacs{PACS numbers:}
%\maketitle

%\section{Model}

\begin{figure}[t]
\includegraphics[width=0.65\columnwidth]{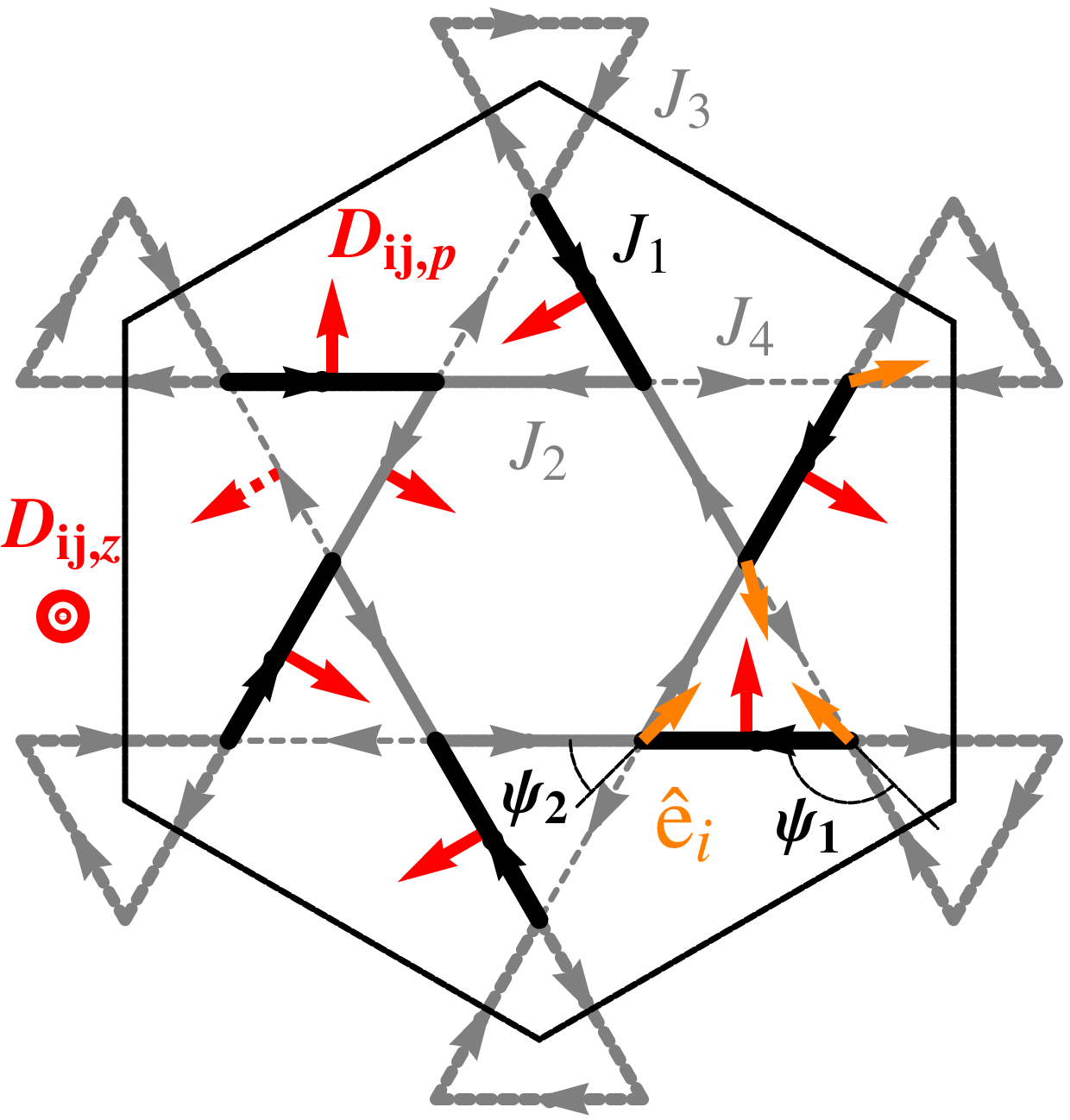}
\caption{Model of DM couplings (red arrows) and transverse magnetic fields (orange arrows) for the pinwheel kagome Rb$_2$Cu$_3$SnF$_{12}$ (not all vectors are shown but they can be deduced by symmetry; $\psi_2 = 51^{\circ}$, $\psi_1 = 131^{\circ}$). The unit-cell (hexagon) is also shown.}
\label{Fig1sm}\end{figure}

We focus on the anticrossing observed in the field-dependence of
the spin gap measured in Rb$_2$Cu$_3$SnF$_{12}$ and describe it by using a model that includes the transverse
anisotropies expected in this compound.\\
\indent In systems where the magnetization $S^z$ is a conserved quantity, the energy gap between the ground state $|S^z=0\rangle$ and the first
triplet state $|S^z=1\rangle$ decreases when an external magnetic
field is applied along the $z$-axis and is expected to vanish once a critical field is reached. This triggers a phase
transition that can be described as a Bose-Einstein condensation of
magnons~\cite{Giamarchi}. In real systems, weak transverse anisotropies of
spin-orbit origin may mix the states with different $S^z$ and lift
the degeneracy at the crossing point, resulting in an anticrossing of
the energy levels and preventing a true condensation. We show here that in the present case this may occur or not, depending on the symmetry of the lowest triplet
state.\\
\indent We consider 
a model previously introduced for Rb$_2$Cu$_3$SnF$_{12}$,
with Heisenberg couplings
and Dzyaloshinskii-Moriya (DM) interactions~\cite{Tanaka,Kim} 
\begin{equation}
{\mathcal H}_0 = \sum_{<i,j>} J_{ij} \mathbf{S}_i \cdot \mathbf{S}_j+ \mathbf{D}_{ij} \cdot (\mathbf{S}_i \times \mathbf{S}_j) -gH \sum_i S^z_i
\end{equation}
where $\mathbf{S}_i$ is the $S=1/2$ operator of the Cu$^{2+}$ spin,
$H$ the external magnetic field along the $z$ direction (along $c$ axis, perpendicular
to the kagome planes), and $g \equiv g_{cc} = 2.44$ the measured gyromagnetic
factor. We recall that there are four inequivalent bonds in the
12-spin unit-cell, described by $J_{ij}=J_{n=1,...,4}$ (see
Fig.~\ref{Fig1sm}). DM vectors are defined on oriented bonds in
Fig.~\ref{Fig1sm}, and characterized by two amplitudes $d_z=D_{n,z}/J_n$
and $d_p=D_{n,p}/J_n$ ($n=1,...,4$) for the out-of-plane and in-plane
components, respectively. The whole pattern of DM vectors can be
deduced by using the C$_3$ rotation and the inversion at the middle of
the pinwheel. As in previous studies, we also neglect the DM component
along the bond, since the lattice distortions allowing for it remain
weak in the Rb compound.\\ 
\indent In addition, there are transverse magnetic fields (with staggered components) within the plane, generated by the off-diagonal elements $g_{z\perp}$ of the $g$ tensor, corresponding to the tilts of the CuF$_6$ octahedra that define the principal axes of the $g$ tensor (see the inset of Fig.~4). These terms can be included in the Hamiltonian as:
\begin{equation}
{\mathcal H} = {\mathcal H}_0 -  g_s \hspace{.05cm} H \sum_{i} \mathbf{\hat{e}}_i \cdot \mathbf{S}_i ,
\end{equation}
where $g_s=0.135$ and the unit-vectors $\mathbf{\hat{e}}_i$ point in the direction of the tilt of the CuF$_6$ octahedra. They are determined by the crystal structure and shown for each site in Fig.~\ref{Fig1sm}, up to symmetry operations. $g_s$ is the off-diagonal tensor component $g_s \equiv g_{z\perp}  = (g_{\parallel} - g_{\perp}) \sin(2 \vartheta)/2$, determined by the anisotropy of the principal-axis values, $g_{\parallel}$ and $g_{\perp}$, of the tensor (supposed to be axial, as is the quadrupolar tensor we determined by NMR), and by the tilt of the principal axis ($\vartheta = 23^{\circ}$) with respect to the crystalline $c$ axis. The estimate of this tilt angle comes from the crystal structure (tilt of the average coordination plane of 4 fluorine first neighbors; see the inset of Fig. 4) and from the corresponding tilt of the quadrupolar tensor, as explained in the manuscript. The experimental values $g_{cc}$ = 2.44 and $g_{aa}$ = 2.15, determined in Ref.~\cite{Morita}, correspond to the rotated $g$ tensor:
\begin{eqnarray}\label{rot}
g_{cc} = \cos^2(\vartheta)g_{\parallel} + \sin^2(\vartheta)g_{\perp} ,\;\;\;\;\, \nonumber \\
g_{aa} = g_{\perp}/2 + [\sin^2(\vartheta)g_{\parallel} + \cos^2(\vartheta)g_{\perp}]/2 ,
\end{eqnarray}
where $g_{aa}$ is the average over all of the in-plane ($\phi$) directions. Inverting equations \ref{rot} leads to $g_{\parallel}$ = 2.497 and $g_{\perp}$ = 2.121. Thus, we obtain an anisotropy of 15\,\% for the $g$ tensor, which characterizes the Cu$^{2+}$ ion coordinated by 4 oxygen or fluorine ions (e.g., see \cite{g}). The relatively large $g_s$ value is then determined by the strong tilt angle $\vartheta$.\\
\indent It is important to observe that neither the in-plane DM interaction nor the transverse fields conserve the magnetization along the $z$-axis and, therefore, both may account for the anticrossing.\\
%\begin{figure}[t]
%\includegraphics[width=0.65\columnwidth]{Fig1SM.eps}
%\caption{Model of DM couplings (red arrows) and transverse magnetic fields (orange arrows) for the pinwheel kagome Rb$_2$Cu$_3$SnF$_{12}$ (not all vectors are shown but they can be deduced by symmetry; $\psi_2 = 51^o$, $\psi_1 = 131^o$). The unit-cell is also shown.}
%\label{Fig1sm}\end{figure} %\section{Perturbation theory}
\indent Perturbation theory in the inter-dimer couplings and non-$S^z$-conserving
terms gives some insights into the properties of the low-energy states. At first-order, the
ground state $|0\rangle$ is a product of singlets on the strongest
bonds (coupling $J_1$ -- there are six such bonds per unit cell) and
the first excited state $|1\rangle$ is a $Q=0$ state with one strong
bond promoted to a triplet~\cite{Tanaka,Kim}. The latter wave functions can be
classified according to the irreducible representations of C$_{3}$ rotation,
$A_{g,u}$ and $E_{g,u}$, where $g$ and $u$ denote even and odd
wave-functions with respect to the inversion at the center of the
pinwheel.  $d_z$ lifts the two-fold degeneracy of the $E_{g,u}$ modes into
$E_{g,u}^{\pm}$ (corresponding to a $\pm 2\pi/3$ phase acquired upon $2\pi/3$ rotation), both
having $S^z=\pm 1$, while the $S^z=0$ states remain unchanged~\cite{Tanaka}.
Changing the sign of $d_z$ interchanges the $+$ and $-$ modes.\\
\indent The level mixing is given by the matrix element $\langle 1|{\mathcal
  H}|0 \rangle$. The state ${\mathcal H}|0 \rangle$ is even under the
inversion and picks up a $-2\pi/3$ phase under rotation, as imposed by
the clockwise rotation pattern of the $\mathbf{D}_{ij,p}$ vectors,
i.e. it transforms as $E_g^-$.  Therefore the matrix element vanishes for all
triplet states $|1\rangle$ except the one with $E_g^-$ symmetry. $E_g$
triplets are indeed predicted to be the two lowest energy states, with
the splitting between $E_g^+$ and $E_g^-$ controlled by $d_z$~\cite{Tanaka,Kim}. For $d_z>0$
(resp. $d_z<0$), the lowest triplet is that with $E_g^-$ symmetry
(resp. $E_g^+$) and there will be, therefore, an anticrossing
(resp. a crossing).  The experimental observation of an anticrossing
is, therefore, only compatible with $d_z>0$ (with the definition of
Fig.~\ref{Fig1sm}). In other words, the lowest triplet state must have
an $E_g^-$ symmetry. This prediction, based on symmetry arguments, is
expected to remain valid beyond perturbation theory, as we shall
confirm next. Interestingly, it can be tested by other experimental
probes, such as polarized neutron inelastic scattering or direct
``forbidden'' optical transitions between the ground state and the
first excited state, as in SrCu$_2$(BO$_3$)$_2$~\cite{cepas}.\\
\indent %\section{Exact diagonalization}
In order to test this prediction beyond perturbation theory and
compute the gap in the relevant limit of strong interdimer couplings,
we have computed the energy gap by exact diagonalizations of
${\mathcal H}$ for small clusters ($N=12,24$ spins). We have fixed the
exchange couplings as previously extracted from neutron scattering
measurements, $J_2=0.95J_1$, $J_3=0.85J_1$, $J_4=0.55J_1$ and
$|d_z|=0.18$ ($J_1=216$~K)~\cite{Tanaka}.\\
\begin{figure}[t]
\includegraphics[width=\columnwidth]{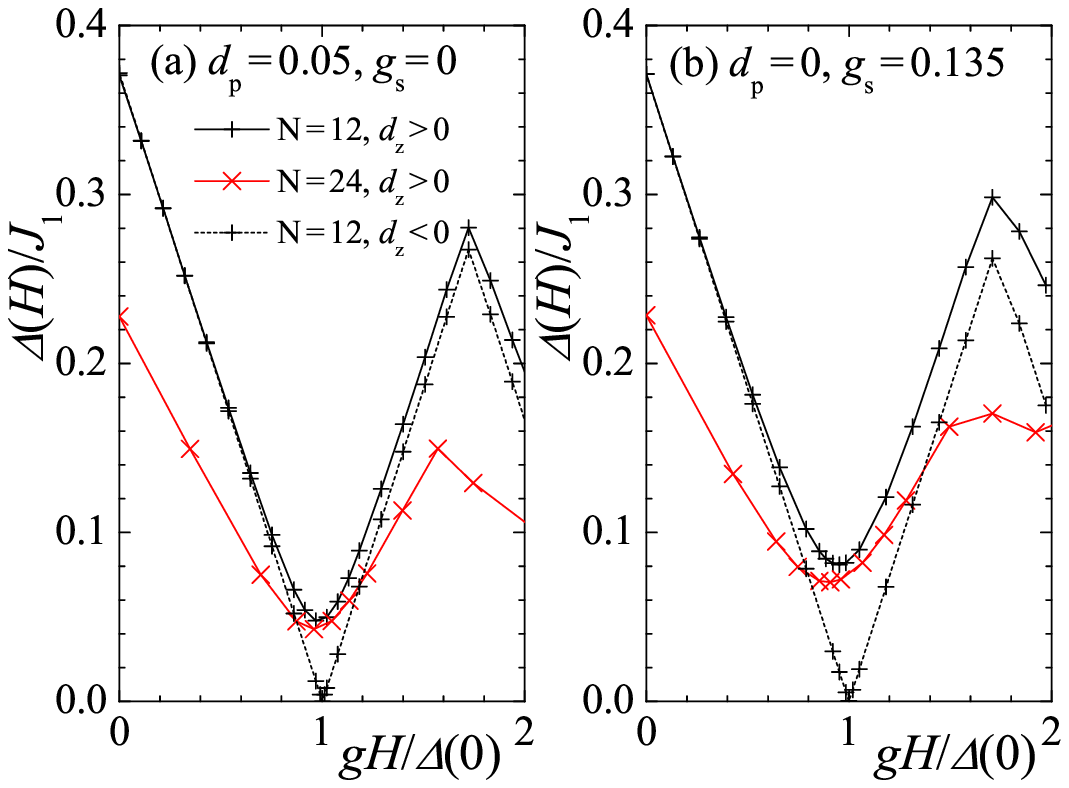}
\caption{Magnetic field dependence of the energy gap, $\Delta(H)$, calculated by exact diagonalization of the Hamiltonian, including transverse DM interactions (a) or transverse fields (b). The anticrossing (resp. crossing) is found for $d_z>0$ (resp. $d_z<0$), as explained by symmetry arguments.}
\label{Fig2sm}
\end{figure}
\indent We show in Fig.~\ref{Fig2sm}(a) the energy gap as a function of the
field for finite transverse DM interaction. First, we find an
anticrossing if $d_z>0$, and a crossing if $d_z<0$, confirming the
symmetry argument given above.  Second, for $d_z>0$, the minimal value
$\Delta_m$ of the gap at the anticrossing increases linearly,
$\Delta_m/J_1 = 0.86 d_p$ at small $d_p$.  It is interesting to
note that finite-size effects on $\Delta_m$ are small compared to
those on $\Delta(0)$ and therefore allows to estimate the anisotropic
coupling. By using $\Delta_m=13$~K, as determined by our NMR study, and $J_1=216$~K, and assuming the
anticrossing would be caused by transverse DM interactions only, we
would have $d_p= 0.07$.\\
\indent However, as mentioned earlier, the staggered fields associated with the off-diagonal component of the $g$ tensor $g_s$ are also expected to
produce an anticrossing (Fig.~\ref{Fig2sm}(b)). This occurs again only
for $d_z>0$ for the same reasons as before.  For $N=24$, and in the
absence of in-plane DM components ($d_p=0$), we find that
$\Delta_m/J_1=0.522 g_s$. As seen in Fig.~\ref{Fig2sm}(b), $\Delta_m$ tends to decrease when increasing the system size.
A crude estimate of the finite-size effects based on the scaling
$\Delta_{m}=\Delta_m^{\infty}+a/N $, leads to 
$\Delta_m^{\infty}/J_1=0.445 g_s$. This gives $\Delta_m^{\infty}=13.0$~K (for $g_s=0.135$) in excellent agreement with the experimental result of $13$~K.
Furthermore, we can also perform the same finite-size scaling on the positions of the minima in Fig.~\ref{Fig2sm}(b), which leads to $g H_m^{\infty} / \Delta(0) = 0.845$. From the experimental value $\Delta(0)$~= 27.2~K we thus find $H_m^{\infty}$~= 14.0~T, again in very good agreement with the experimental value of 13.3(5)~T (see Fig.~4).\\
\indent These estimates show that the strong tilting of the CuF$_6$ octahedra is itself sufficient to \textit{quantitatively} account for the observed residual gap. If any in-plane component of the DM interaction exists, it should provide only a minor effect on the final size of $\Delta_m$. Note that we cannot exclude that the $N=24$ gap has already converged; the corresponding estimate of $d_p$ value will then provide the upper bound. Assuming that the two effects are additive, as they are in the lowest order, $\Delta_m/J_1=|0.522 g_s + 0.86 d_p|$ for $N=24$, we find that $|d_p| < 0.012$, possibly compatible with zero. In a similar way (using $N \rightarrow \infty$ expressions), a reasonable estimate for the maximum error of $\pm 15 \%$ in the determination of the $g_s$ value leads to the same constraint $\left|d_p\right| < 0.012$.\\
\indent We therefore conclude that the anticrossing is primarily caused by the $g$-tensor tilts, with in-plane DM interaction terms ($|d_p|<0.012$) providing a minor correction to this effect. The anticrossing appears only if the helicity of the $g$-tensor tilts (here defined in Fig.~\ref{Fig1sm}) are the same as that defined by the screw axis $d_z$, that is if  $d_z>0$. It is interesting to note that the same $g$-tensor tilts with $d_z<0$ would produce a (previously expected) level-crossing, and thus a sharp transition at the expected critical field value $H_c=\Delta(0)/g$.


\begin{thebibliography}{25}
\bibitem{comment1}
The spectral lines could not have been observed any more at higher temperatures because the short spin spin correlation time ($T_2$) strongly reduces the signal intensity.
\bibitem{comment2}
All central NMR lines follow the same temperature dependence, meaning that the gap opens homogeneously in the system.
\end{thebibliography}
\end{document}